\documentclass[letter]{ieice}
\usepackage[pdftex]{graphicx,xcolor}%
\usepackage[fleqn]{amsmath}
\usepackage{newtxtext}
\usepackage[varg]{newtxmath}
\usepackage[square,sort,comma,numbers]{natbib}
\usepackage{color}
\usepackage{colortbl}
\usepackage{multirow}
\usepackage{algorithmic}
\usepackage{graphicx}
\usepackage{textcomp}
\usepackage{tcolorbox}
\usepackage{multirow}
\usepackage{url}
\usepackage{listings}
\usepackage{comment}
\usepackage{lscape}
\usepackage{CJK}
\usepackage{balance}
\usepackage{subfigure}
\usepackage{float}

\newcommand{\AmSLaTeX}{%
 $\mathcal A$\lower.4ex\hbox{$\!\mathcal M\!$}$\mathcal S$-\LaTeX}
\def\BibTeX{{\rmfamily B\kern-.05em
 \textsc{i\kern-.025em b}\kern-.08em
  T\kern-.1667em\lower.7ex\hbox{E}\kern-.125emX}}
\makeatletter
\def\tmpcite#1{\@ifundefined{b@#1}{\textbf{?}}{\csname b@#1\endcsname}}%
\makeatother

\field{D}
\vol{105}
\no{1}
\SpecialSection{Empirical Software Engineering}
\title
      {\tool[]: Function-Call Reachability Detection of Vulnerable Code for \texttt{npm} Packages}
\authorlist{%
\authorentry[bodin.chinthanet.ay1@is.naist.jp]{Bodin Chinthanet}{n}{naist}\MembershipNumber{}
\authorentry[raula-k@is.naist.jp]{Raula Gaikovina Kula}{n}{naist}\MembershipNumber{}
\authorentry[rodrigo.elizalde.qy5@is.naist.jp]{Rodrigo Eliza Zapata}{n}{naist}\MembershipNumber{}
\authorentry[ishio@is.naist.jp]{Takashi Ishio}{m}{naist}\MembershipNumber{1301152}
\authorentry[matumoto@is.naist.jp]{Kenichi Matsumoto}{f}{naist}\MembershipNumber{8925358}
\authorentry[ihara@wakayama-u.ac.jp]{Akinori Ihara}{m}{wakayama}\MembershipNumber{0728160}
}
\affiliate[naist]{The author is with Nara Institute of Science and Technology, Nara, Japan}
\affiliate[wakayama]{The author is with Wakayama University, Wakayama, Japan}

\usepackage[colorlinks = true,
            linkcolor = blue,
            urlcolor  = blue,
            citecolor = blue,
            anchorcolor = blue]{hyperref}

\definecolor{codegreen}{rgb}{0,0.6,0}
\definecolor{codegray}{rgb}{0.5,0.5,0.5}
\definecolor{codepurple}{rgb}{0.58,0,0.82}
\definecolor{backcolour}{rgb}{0.95,0.95,0.92}

\lstdefinestyle{mystyle}{
	backgroundcolor=\color{backcolour},   
	commentstyle=\color{codegreen},
	keywordstyle=\color{magenta},
	numberstyle=\tiny\color{codegray},
	stringstyle=\color{codepurple},
	basicstyle=\footnotesize,
	breakatwhitespace=false,         
	breaklines=true,                 
	captionpos=b,                    
	keepspaces=true,                 
	numbers=left,                    
	numbersep=2pt,                  
	showspaces=false,                
	showstringspaces=false,
	showtabs=false,                  
	tabsize=2
}

\newcommand{\tool}[1][]{S\={o}jiTantei}

\newcommand\revision[1]{#1}

\begin{document}
\maketitle

\begin{summary}
It has become common practice for software projects to adopt third-party dependencies.
Developers are encouraged to update any outdated dependency to remain safe from potential threats of vulnerabilities. 
In this study, we present an approach to aid developers show whether or not a vulnerable code is reachable for JavaScript projects.
Our prototype, \tool[], is evaluated in two ways (i) the accuracy when compared to a manual approach and (ii) a larger-scale analysis of 780 clients from 78 security vulnerability cases.
The first evaluation shows that \tool[] has a high accuracy of 83.3\%, with a speed of less than a second analysis per client. 
The second evaluation reveals that 68 out of the studied 78 vulnerabilities reported having at least one clean client.
The study proves that automation is promising with the potential for further improvement.
\end{summary}

\section{Introduction}
Raising the awareness for developers to quickly update their third-party dependencies is now regarded as the priority \citep{Kula:2018, Kikas:2017}, especially if the threat includes malicious intent.
As well as fixing bugs and adding new features, migration to a new version (i.e., update) sometimes includes fixes to prevent these threats.
Such threats on dependency are regarded as \textit{vulnerable dependency}.
Recent epidemic vulnerabilities such as \texttt{heartbleed} \citep{Web:heartbleed} are examples of how vulnerable dependencies can affect all users in an ecosystem of users such as the \texttt{npm} ecosystem.

Recent studies \citep{Hejderup:2018,Ponta:EMSE2020} have supported the claim of overestimating vulnerability alerts, where the client does not actually call the vulnerable code but specifically for the Java programming language.
\citet{Ponta:EMSE2020} showed that, for \texttt{Java} projects, many clients do not actually call the affected function.
We refer to these projects as being \textit{clean} (i.e., they do not execute the affecting code in their client applications).
Conversely, we refer to \textit{reached} clients as projects that adopt and execute the vulnerability code.
Because there are many static analysis tools for the Java programming language, such code-centric is possible. However, as stated by \citet{Chinthanet:ASE2020}, this is not the case for programming languages like JavaScript and npm packages.

In our prior work \citep{zapata2018towards}, we show that there is indeed an overestimation, with 73.3\% of outdated clients were actually not in direct danger from the threat.
In this work, we analyze the effect of vulnerabilities within the npm ecosystem on a larger scale than the manual study of 60 clients.
\revision{We develop a tool called \tool~that parses JavaScript codes in the client application to find the lines of code that call the vulnerable function from dependencies.}
Our tool was evaluated in two experiments: (i) a replication study for accuracy and (ii) a larger-scale analysis of vulnerabilities.
The replication dataset for our work is available at \url{https://github.com/NAIST-SE/SojiTantei}

\section{Experiments}
\label{sec:evaluation}

We carried out two experiments.
The first experiment was a replication study of \citet{zapata2018towards} by using \tool.
We show a comparison of our results against this manual work.

For the second experiment, our aim is to analyze a larger statistical sample set of projects.
We draw from the \citet{decan2018impact} study, using a stratified sample from the 400 vulnerabilities. (with a confidence level of 95\% and a confidence interval of 5.\footnote{\url{https://www.surveysystem.com/sscalc.htm}})
The final dataset that matched our criteria included 780 clients affected by 78 vulnerabilities, ending with 196 vulnerabilities.
To ensure a quality dataset, we then selected vulnerabilities that met the following criteria: (i) were accessible to be downloaded, (ii) had at least 10 clients (i.e., similar to the prior study) that we could test, and (iii) the vulnerability had to have a fix in which, we could identify the vulnerable function.
It is important to note that the detection of the vulnerable function was still performed manually.
For our results, we will report the proportion of projects using the vulnerable function.
Note that listed dependencies refer to clients that list the vulnerable dependency, but the client code does not call any functions of the dependency.
Furthermore, we also investigate whether the vulnerable dependency was updated or not (i.e., for the \tool[] performance).
\begin{figure}[t]
\centering
\includegraphics[width=.5\textwidth]{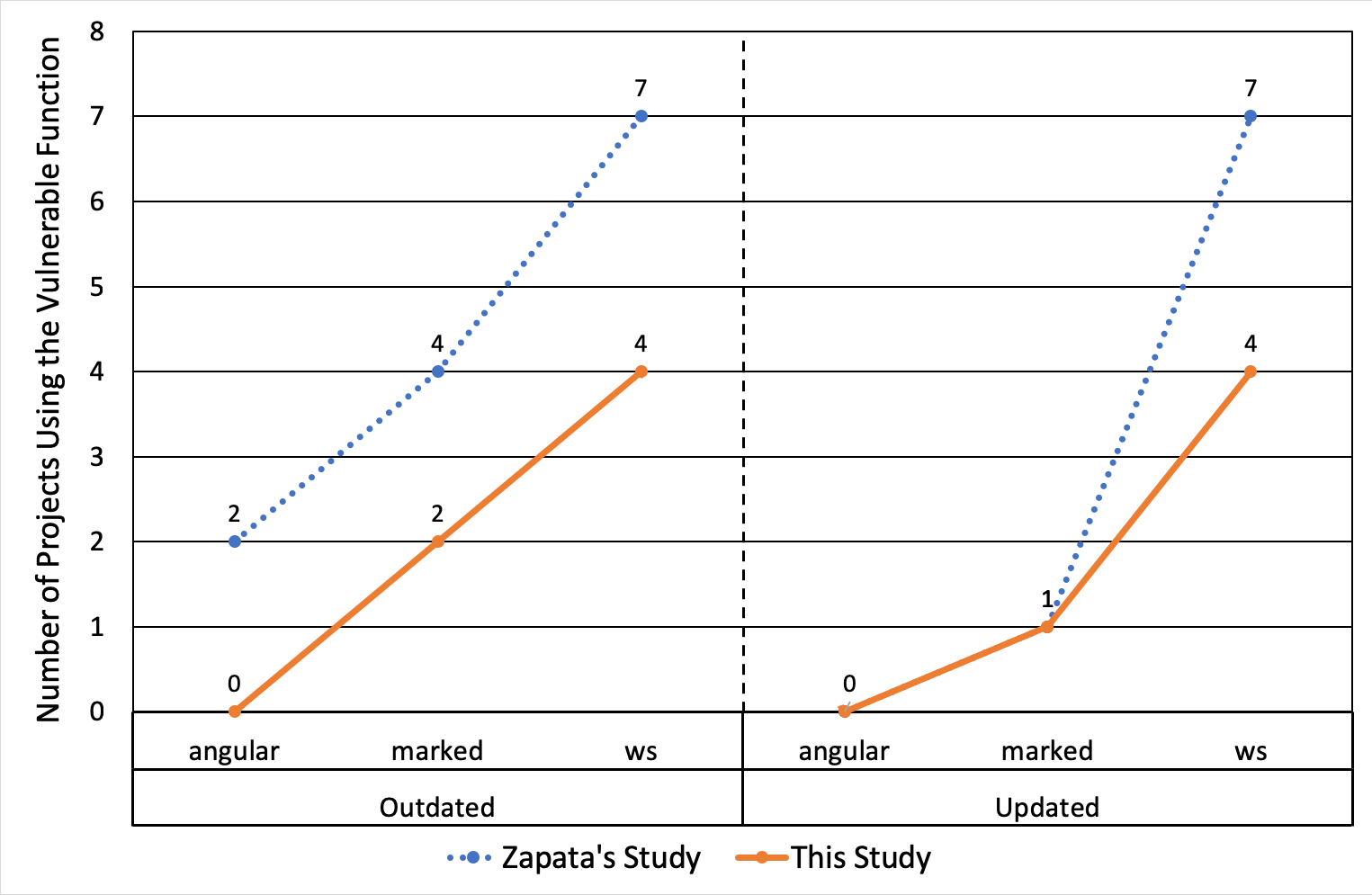}
\caption{Comparison of results between the study of \citeauthor{zapata2018towards} and this study. Results show that our method does not capture all vulnerable projects.}
\label{fig:replicationResults}
\end{figure}

\begin{table}[t]
\centering
\caption{Results of Performance Metrics }
\label{tab:replication_statistics}
\scalebox{1}{
\begin{tabular}{|l|c|l|}
\hline
\multicolumn{1}{|c|}{\textbf{Metric}} & 
\multicolumn{1}{c|}{\textbf{Value}} & 
\multicolumn{1}{c|}{\textbf{Description}} \\ \hline
\multicolumn{3}{|c|}{\textbf{Statistical Measures}} \\ \hline
n & 60 & \# projects \\ \hline
TN & 39 & True Negatives. \\ \hline
FN & 10 & False Negatives.\\ \hline
FP & 0 & False Positives. \\ \hline
TP & 11 & True Positives. \\ \hline
\multicolumn{3}{|c|}{\textbf{Performance Measures}} \\ \hline
Accuracy & \cellcolor{blue!25} 0.833 & (TN + TP)/N\\ \hline
Miss-classification rate & 0.167 & (FP + FN)/N\\ \hline
TP rate & 0.524 & TP/(FN + TP)\\ \hline
FP rate & 0 & FP/(TN + FP)\\ \hline
TN rate & 1 & TN/(TN + FP)\\ \hline
FN rate & 0.476 & FN/(FN + TP)\\ \hline
\end{tabular}
}
\end{table}
\textbf{Results for \tool[] performance}.
Table \ref{tab:replication_statistics} presents results for the first experiment.
Importantly, we find that \tool[] has an accuracy of 83.3\% out of the 60 projects that were studied. 
This indicates that our method is reliable for the sample projects.
As shown in Figure \ref{fig:replicationResults}, \tool[] is not perfect, as not report all detected functions (i.e., these are reported as 10 false negatives).
\revision{This false negative is due to \tool[] cannot capture the function call written in the new standard (e.g., ES6).}

Additionally, it is important to mention that the average execution time for the function-call extraction was 0.73 seconds per project, making this approach significantly faster than the manual execution.
This is especially significantly faster when compared to a manual analysis task.

\begin{table}[t]
\caption{Summary of client classifications by \tool[]}
\label{tab:clientClassificationResults}
\centering
\scalebox{1}{
\begin{tabular}{|l|l|}
\hline
\multicolumn{1}{|c|}{\textbf{Client Classification}} & 
\multicolumn{1}{c|}{\textbf{\# Clients}} \\ \hline
Clean & 249\\ 
Reached & 33\\ \hline
Listed Only & 445\\ 
No Data Available & 53\\ \hline \hline
Total & 780\\ \hline
\end{tabular}
}
\end{table}

\textbf{Results for the Case Study}
Table \ref{tab:clientClassificationResults} shows the results of the analysis of the output generated by \tool[], finding that 249 of the clients were using the vulnerable package but not actually have a direct function-call to the vulnerable code. 
Overall, we found that most clients (61 vulnerabilities) did not reach the vulnerable code. 
We found that 61 of the vulnerabilities had no clients that were directly using the vulnerable code (0\% reached).
There were vulnerabilities that reached at least one client (i.e., 1\%$\sim$99\%). 
In this case, around a median of 42.86\% of the total clients for each vulnerability had clean clients.

\section{Implications and Future Directions}
\label{sec:conclusion}
We summarize the implications of the study and its contributions as follows: (1) It is likely for the client not to reach the dependency vulnerable code. (2) Automation is promising with the potential for improvement. (3) We believe that knowing that a client is clean can motivate client developers to update dependencies and that the community can contribute to finding accurate methods to report vulnerabilities threats.
\revision{The immediate future are (1) integrating \tool[] with the automated pull request bot such as Dependabot \footnote{\url{https://dependabot.com/}} and (2) extending the study to other programming languages.}

\section*{Acknowledgement}
This work was supported by JSPS KAKENHI Grant Numbers JP18H04094, JP18H03221, 20K19774, 20H05706.

\bibliographystyle{abbrvnat}
\bibliography{bibliography.bib}%
\end{document}